\documentclass[pdflatex,sn-mathphys-num]{sn-jnl}% Math and Physical Sciences Numbered 
\usepackage{graphicx}%
\usepackage{lineno}
\usepackage{multirow}%
\usepackage{amsmath,amssymb,amsfonts}%
\usepackage{amsthm}%
\usepackage{mathrsfs}%
\usepackage{xcolor}%
\usepackage{colortbl}%
\usepackage{textcomp}%
\usepackage{manyfoot}%
\usepackage{booktabs}%
\usepackage{algorithm}%
\usepackage{algorithmicx}%
\usepackage{algpseudocode}%
\usepackage{array}
\usepackage{bm}
\usepackage[caption=false,font=normalsize,labelfont=sf,textfont=sf]{subfig}
\usepackage{stfloats}
\usepackage{url}
\usepackage{verbatim}
\usepackage{diagbox}
\newcommand{\tabincell}[2]{\begin{tabular}{@{}#1@{}}#2\end{tabular}}
\usepackage{hyperref}
\hypersetup{hypertexnames=false}

\theoremstyle{thmstyleone}

\theoremstyle{thmstyletwo}%

\theoremstyle{thmstylethree}%

\begin{document}
\title[Real-time quasi-global precipitation retrieval]{Huayu-Global: Skillful real-time quasi-global precipitation retrieval} %% Article title

\author[1,2]{Zijiang~Song}
\email{jacksung1995@gmail.com}
\author[1,2]{Shuang~Yang}
\author[1,2]{Yuying~Li}
\author[1,2]{Zilong~Zhu}
\author[3]{Xigang~Sun}
\author[1,2]{Ting~Liu}
\author[1,2]{Wenye~Xiao}
% \author[1,2]{Ao~Xu}
% \author[1,2]{Ye~Li\corref{cor1}}
% \ead{yli@geo.ecnu.edu.cn}
% \author[4]{Feng~Lu}
\author*[1,2]{Lina~Yuan}
\email{lnyuan@geoai.ecnu.edu.cn}
\author*[1,2]{Min~Liu}
\email{mliu@geo.ecnu.edu.cn}

% \author[1,2]{\fnm{Zijiang} \sur{Song}}\email{zjsong@stu.ecnu.edu.cn}
% \author[1,2]{\fnm{Ting} \sur{Liu}}
% \author*[1,2]{\fnm{Lina} \sur{Yuan}}\email{lnyuan@geoai.ecnu.edu.cn}
% \author[1,2]{\fnm{Yuying} \sur{Li}}
% \author[1,2]{\fnm{Ao} \sur{Xu}}
% \author[3]{\fnm{Xigang} \sur{Sun}}
% \author*[1,2]{\fnm{Ye} \sur{Li}}\email{yli@geo.ecnu.edu.cn}
% \author[4]{\fnm{Feng} \sur{Lu}}
% \author*[1,2]{\fnm{Min} \sur{Liu}}\email{mliu@geo.ecnu.edu.cn}

\affil[1]{\orgdiv{Key Laboratory of Geographic Information Science (Ministry of Education)}, \orgname{East China Normal University}, \orgaddress{\city{Shanghai}, \postcode{200241}, \country{China}}}

\affil[2]{\orgdiv{Key Laboratory of Spatial-temporal Big Data Analysis and Application of Natural Resources in Megacities (Ministry of Natural Resources)}, \orgname{East China Normal University}, \orgaddress{\city{Shanghai}, \postcode{200241}, \country{China}}}

\affil[3]{\orgdiv{School of Computer Science and Engineering}, \orgname{Southeast University}, \orgaddress{\city{Nanjing}, \postcode{211189}, \country{China}}}

% \affil[4]{\orgname{Key Laboratory of Radiometric Calibration and Validation for Environmental Satellites, National Satellite Meteorological Center (National Center for Space Weather), Innovation Center for Feng Yun Meteorological Satellite (FYSIC), China Meteorological Administration}, \orgaddress{\city{Beijing}, \postcode{100081}, \country{China}}}

% \linenumbers
\abstract{
Satellite precipitation products face a persistent trade-off: real-time estimates are timely but less accurate, whereas the most reliable products are often released months after observation. This delay keeps high-quality precipitation information outside the decision window for flood warning and disaster response. Here we report Huayu-Global, a cross-constellation deep-learning system that combines three geostationary sensor streams from five satellites through a common retrieval framework and generates coherent quasi-global precipitation fields every thirty minutes. Across more than a million independent station records from two periods outside the training years, Huayu-Global improves event detection and precipitation-amount estimation relative to the real-time products evaluated and reaches delayed-product-level skill in key metrics. Huayu-Global therefore turns operational geostationary observations into a timely, high-skill precipitation layer for flood forecasting, water management and disaster response. Its cross-constellation design also provides a scalable foundation for incorporating future geostationary missions into hydrological and weather-prediction systems.
}

\maketitle
\section{Introduction}
\begin{figure}[!htbp]
\centering
\includegraphics[width=0.85\textwidth]{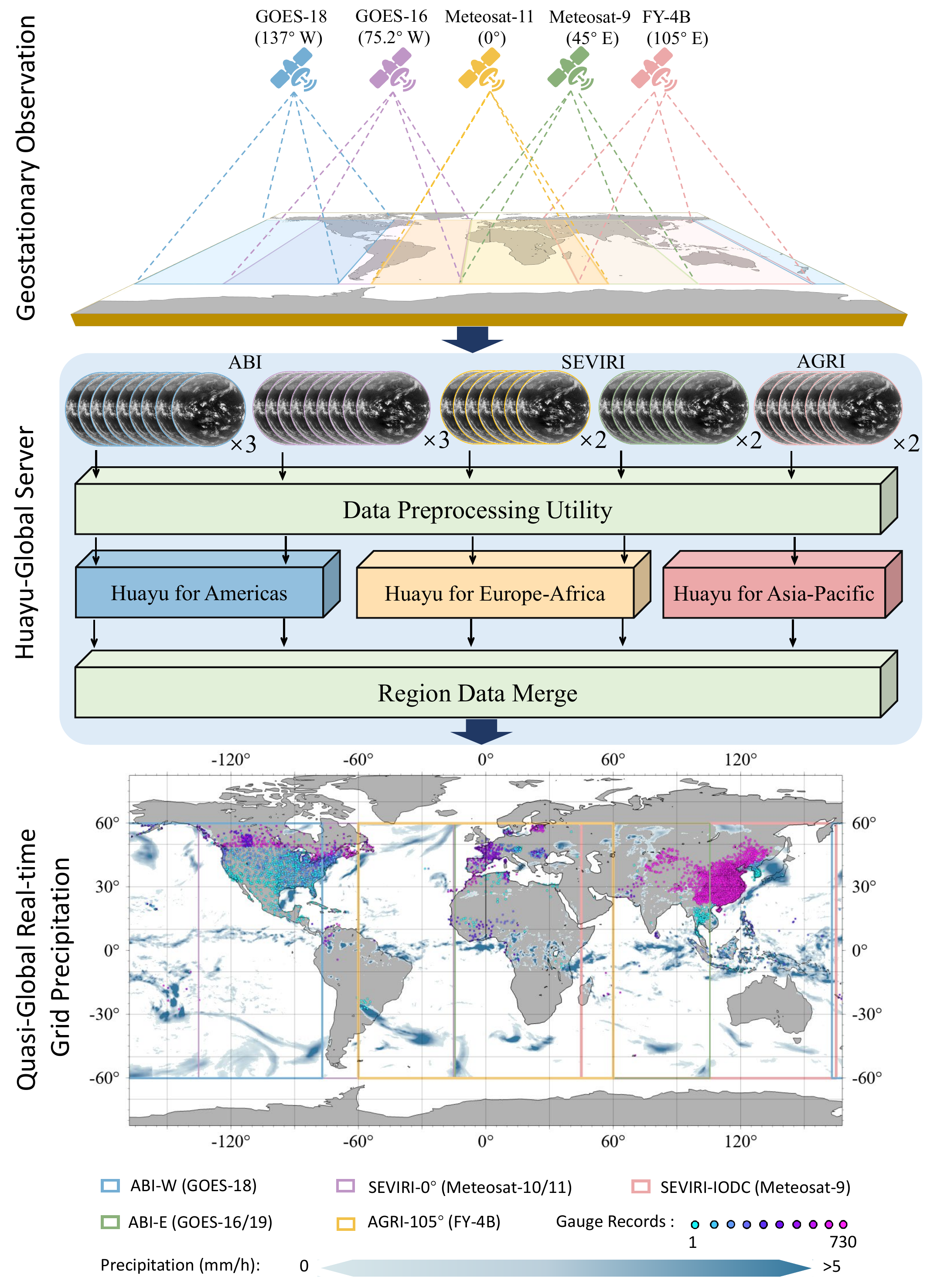}
\caption{Data processing workflow for the Huayu-Global precipitation monitoring framework. Representative Geostationary Operational Environmental Satellite (GOES), Meteosat, and FengYun-4B (FY-4B) sector views are transformed into Advanced Baseline Imager (ABI), Spinning Enhanced Visible and Infrared Imager (SEVIRI), and Advanced Geostationary Radiation Imager (AGRI) cloud-feature sequences, then passed through the Huayu models for the Americas, Europe-Africa, and Asia-Pacific domains. Sector outputs are merged to generate coherent quasi-global grid precipitation products using only satellite inputs available at the inference time.}
\label{fig:processing}
\end{figure}

Precipitation is a primary driver of the water cycle, flood hazard, agricultural risk, and hydrological extremes. Operational monitoring therefore requires products that are both timely and quantitatively reliable. Ground gauges, weather radar, and gauge--satellite merged analyses provide essential constraints on precipitation fields, but their global operational use is limited by uneven station density, radar-network gaps, terrain blockage, oceanic coverage gaps, and differences in data-sharing conditions~\citep{shen2014high,fulton1998wsr,sun2018review,kirschbaum2017nasa}. These limitations are most acute when precipitation information is needed rapidly and consistently across continents and ocean-adjacent regions.

Satellite precipitation products have reduced this coverage gap, but operational users still face a hard choice between speed and accuracy. Real-time streams such as Global Satellite Mapping of Precipitation NOW (GSMaP\_NOW) and Precipitation Estimation from Remotely Sensed Information using Artificial Neural Networks--Cloud Classification System (PERSIANN-CCS) provide prompt coverage but show lower detection and quantitative skill (real-time products achieve critical success index (CSI) values between 0.23 and 0.33 in our evaluation, whereas the gauge-adjusted Final Run of Integrated Multi-satellitE Retrievals for Global Precipitation Measurement (IMERG FR), released 3.5~months after observation, reaches 0.42)~\citep{kubota2020global,nguyen2018persiann,hong2004precipitation}. More extensively processed products such as the Climate Prediction Center morphing technique (CMORPH) and IMERG can improve accuracy, but their highest-quality streams may arrive months later~\citep{joyce2004cmorph,huffman2015nasa,kidd2020global,tan2019imerg,huffman2023imerg}. Global IMERG assessments also show that residual errors remain dependent on region, season, and precipitation regime, even after multi-sensor merging and gauge adjustment~\citep{tan2016novel,pradhan2022review}. Information delivered after the decision window cannot support flood warning or other time-critical action. A useful quasi-global system must therefore combine real-time availability with delayed-product-level precipitation skill.

High-frequency geostationary observations offer a route to this goal because they continuously sample cloud evolution across broad domains. Decades of geostationary meteorological satellite development, from early operational systems to modern multispectral imagers, have increased the temporal and spectral information available for rainfall inference~\citep{blersch1991geostationary,schmit2005introducing,bessho2016introduction}. The difficulty is that cloud-top radiances do not map uniquely to surface rainfall, and the relationship varies with region, season, terrain, storm type, and sensor geometry. The challenge is compounded across satellite platforms: the GOES ABI, Meteosat SEVIRI, and FY-4B AGRI differ in spectral coverage, native spatial resolution, and full-disk scan cadence, yet an operational quasi-global product must seamlessly combine their outputs. Recent studies have shown that nonlinear and multispectral models can improve geostationary precipitation retrieval in regional settings~\citep{song2026huayu,kuhnlein2014improving,sehad2017novel,wang2020infrared,wang2021infrared,ma2022fy4qpe,ma2022improvement,zhang2022precipgradenet,zhu2024pecam,ji2025new}. The remaining challenge is to extend this idea from single-sensor or regional retrievals to a cross-constellation system that can operate across the Americas, Europe-Africa, and Asia-Pacific sectors while preserving a common output grid.

Here we present Huayu-Global, a cross-constellation framework for real-time quasi-global precipitation retrieval. Extending the earlier regional Huayu framework~\citep{song2026huayu}, it coordinates sensor-adapted models for GOES ABI, Meteosat SEVIRI, and FY-4B AGRI within a common architecture and merges their outputs on a unified 0.05$^\circ$ grid (Fig.~\ref{fig:processing}). Huayu-Global thereby converts operational geostationary observations into coherent precipitation fields every 30~minutes across most low- and mid-latitude regions. The system is developed from 59.7~TB of satellite data and evaluated against 5,395 independent ground stations contributing 1.85~million matched three-hourly records from July--December~2022 and January--June~2025, both outside the 2023--2024 training years. It leads the evaluated real-time products in event detection and error magnitude and reaches delayed-product-level skill in key station-based metrics. By moving high-skill precipitation information from retrospective archives into the operational decision window, Huayu-Global establishes a scalable observation layer for flood forecasting, water management, disaster response and future satellite-enabled weather prediction.

\section{Results}

\subsection{Availability--skill comparison}

\begin{figure}[!htbp]
\centering
\includegraphics[width=0.88\textwidth]{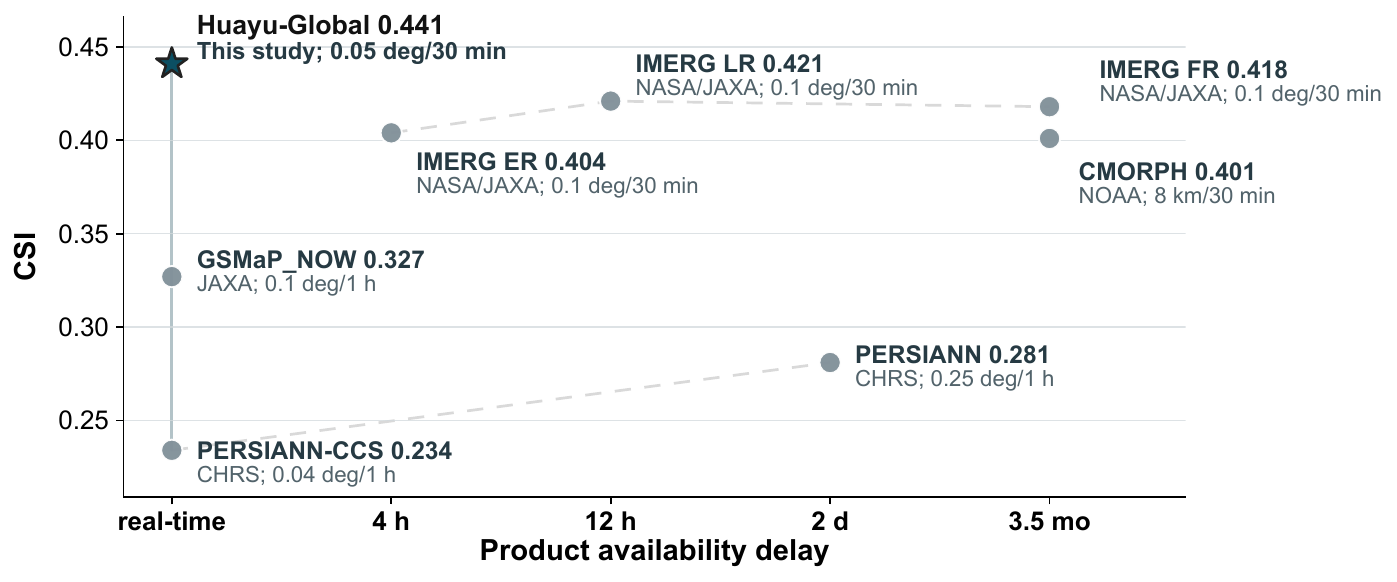}
\caption{Detection skill by product availability class. The vertical axis shows the full-sample CSI from the independent station evaluation, and the horizontal axis shows documented product-release or processing classes rather than measured end-to-end processing time. Huayu-Global and the real-time baseline products are grouped in the real-time class; IMERG Early Run (IMERG ER), IMERG Late Run (IMERG LR), PERSIANN, CMORPH, and IMERG FR are displayed according to their documented release categories. Lines connect product streams within the same product family where applicable.}
\label{fig:figure_latency}
\end{figure}

Huayu-Global was evaluated against seven satellite precipitation products spanning real-time, short-delay, and delayed post-processed streams. Figure~\ref{fig:figure_latency} summarizes the main availability--skill comparison: within the matched station sample, Huayu-Global is the only real-time product whose critical success index (CSI) is higher than all evaluated baselines, including products released months later.

\begin{table*}[t]
\scriptsize
\setlength{\extrarowheight}{2pt}
\setlength{\tabcolsep}{4pt}
\centering
\caption{Overall performance comparison over 1,847,967 matched 3-hour station-product records from July--December~2022 and January--June~2025. Huayu-Global is compared with seven satellite precipitation products on the same matched station windows using CSI, ACC, R$^2$, CC, and RMSE. All values in this table are evaluated against independent station observations. Smaller values below each estimate denote $\pm$ one-half of the central 95\% bootstrap confidence interval from 1,000 station--date-block resamples (Methods). Bold values indicate the best value in each metric column.}
\label{tab:metrics_selected}
\begin{tabular*}{\textwidth}{@{\extracolsep{\fill}}lccccccc@{}}
\toprule
\tabincell{c}{\textbf{Product}}
&\tabincell{c}{\textbf{Resolution}}
&\tabincell{c}{\textbf{Latency}}
&\tabincell{c}{\textbf{CSI}$\uparrow$}
&\tabincell{c}{\textbf{ACC}$\uparrow$}
&\tabincell{c}{\textbf{R$^2$}$\uparrow$}
&\tabincell{c}{\textbf{CC}$\uparrow$}
&\tabincell{c}{\textbf{RMSE}$\downarrow$\\(mm)}
\\
\midrule
\tabincell{c}{PERSIANN-CCS} & \tabincell{c}{0.04°/1h} & \tabincell{c}{real-time}
& \tabincell{c}{$0.234$\\[-5px]{\fontsize{6}{0} $\pm0.001$}}
& \tabincell{c}{$0.853$\\[-5px]{\fontsize{6}{0} $\pm0.001$}}
& \tabincell{c}{$-0.177$\\[-5px]{\fontsize{6}{0} $\pm0.020$}}
& \tabincell{c}{$0.334$\\[-5px]{\fontsize{6}{0} $\pm0.008$}}
& \tabincell{c}{$2.660$\\[-5px]{\fontsize{6}{0} $\pm0.032$}}\\

\tabincell{c}{GSMaP\_NOW} & \tabincell{c}{0.1°/1h} & \tabincell{c}{real-time}
& \tabincell{c}{$0.327$\\[-5px]{\fontsize{6}{0} $\pm0.001$}}
& \tabincell{c}{$0.859$\\[-5px]{\fontsize{6}{0} $\pm0.001$}}
& \tabincell{c}{$-0.374$\\[-5px]{\fontsize{6}{0} $\pm0.036$}}
& \tabincell{c}{$0.387$\\[-5px]{\fontsize{6}{0} $\pm0.008$}}
& \tabincell{c}{$2.874$\\[-5px]{\fontsize{6}{0} $\pm0.039$}}\\
\midrule
\tabincell{c}{CMORPH} & \tabincell{c}{8km/30min} & \tabincell{c}{3-4 months}
& \tabincell{c}{$0.401$\\[-5px]{\fontsize{6}{0} $\pm0.002$}}
& \tabincell{c}{$\textbf{0.904}$\\[-5px]{\fontsize{6}{0} $\pm0.000$}}
& \tabincell{c}{$0.255$\\[-5px]{\fontsize{6}{0} $\pm0.013$}}
& \tabincell{c}{$0.597$\\[-5px]{\fontsize{6}{0} $\pm0.008$}}
& \tabincell{c}{$2.118$\\[-5px]{\fontsize{6}{0} $\pm0.035$}}\\

\tabincell{c}{PERSIANN} & \tabincell{c}{0.25°/1h} & \tabincell{c}{2 days}
& \tabincell{c}{$0.281$\\[-5px]{\fontsize{6}{0} $\pm0.001$}}
& \tabincell{c}{$0.855$\\[-5px]{\fontsize{6}{0} $\pm0.001$}}
& \tabincell{c}{$0.156$\\[-5px]{\fontsize{6}{0} $\pm0.007$}}
& \tabincell{c}{$0.421$\\[-5px]{\fontsize{6}{0} $\pm0.007$}}
& \tabincell{c}{$2.257$\\[-5px]{\fontsize{6}{0} $\pm0.037$}}\\

\tabincell{c}{IMERG ER} & \tabincell{c}{0.1°/30min} & \tabincell{c}{4 hours}
& \tabincell{c}{$0.404$\\[-5px]{\fontsize{6}{0} $\pm0.002$}}
& \tabincell{c}{$0.876$\\[-5px]{\fontsize{6}{0} $\pm0.000$}}
& \tabincell{c}{$0.195$\\[-5px]{\fontsize{6}{0} $\pm0.013$}}
& \tabincell{c}{$0.553$\\[-5px]{\fontsize{6}{0} $\pm0.008$}}
& \tabincell{c}{$2.203$\\[-5px]{\fontsize{6}{0} $\pm0.034$}}\\

\tabincell{c}{IMERG LR} & \tabincell{c}{0.1°/30min} & \tabincell{c}{12 hours}
& \tabincell{c}{$0.421$\\[-5px]{\fontsize{6}{0} $\pm0.002$}}
& \tabincell{c}{$0.887$\\[-5px]{\fontsize{6}{0} $\pm0.000$}}
& \tabincell{c}{$0.244$\\[-5px]{\fontsize{6}{0} $\pm0.013$}}
& \tabincell{c}{$0.590$\\[-5px]{\fontsize{6}{0} $\pm0.007$}}
& \tabincell{c}{$2.134$\\[-5px]{\fontsize{6}{0} $\pm0.034$}}\\

\tabincell{c}{IMERG FR} & \tabincell{c}{0.1°/30min} & \tabincell{c}{3.5 months}
& \tabincell{c}{$0.418$\\[-5px]{\fontsize{6}{0} $\pm0.002$}}
& \tabincell{c}{$0.885$\\[-5px]{\fontsize{6}{0} $\pm0.000$}}
& \tabincell{c}{$0.307$\\[-5px]{\fontsize{6}{0} $\pm0.012$}}
& \tabincell{c}{$\textbf{0.613}$\\[-5px]{\fontsize{6}{0} $\pm0.007$}}
& \tabincell{c}{$2.042$\\[-5px]{\fontsize{6}{0} $\pm0.035$}}\\

\midrule
\tabincell{c}{\textbf{Huayu-Global}} & \tabincell{c}{0.05°/30min} & \tabincell{c}{real-time}
& \tabincell{c}{$\textbf{0.441}$\\[-5px]{\fontsize{6}{0} $\pm0.002$}}
& \tabincell{c}{$0.896$\\[-5px]{\fontsize{6}{0} $\pm0.000$}}
& \tabincell{c}{$\textbf{0.362}$\\[-5px]{\fontsize{6}{0} $\pm0.008$}}
& \tabincell{c}{$0.605$\\[-5px]{\fontsize{6}{0} $\pm0.007$}}
& \tabincell{c}{$\textbf{1.961}$\\[-5px]{\fontsize{6}{0} $\pm0.038$}}\\

\bottomrule
\end{tabular*}
\end{table*}

The full-sample metrics in Table~\ref{tab:metrics_selected} show the same pattern across complementary diagnostics. Across 1,847,967 matched 3-hour station-product records, Huayu-Global achieves the highest CSI (0.441) and coefficient of determination (R$^2$, 0.362), and the lowest root-mean-square error (RMSE, 1.961~mm). Huayu-Global also exceeds the 3.5-month IMERG FR product in CSI, R$^2$, and RMSE in this independent station comparison. Its overall accuracy (ACC, 0.896) is close to the highest value from CMORPH (0.904), and its correlation coefficient (CC, 0.605) is close to the highest value from IMERG FR (0.613). Daily-scale diagnostics (Extended Data Fig.~\ref{fig:figure_daily_combined}) show that this aggregate advantage is not confined to a small number of validation dates, while also indicating that the strongest non-Huayu baseline varies by metric and day.

\subsection{Representative spatial case}

\begin{figure}[!htbp]
\centering
\includegraphics[width=0.98\textwidth]{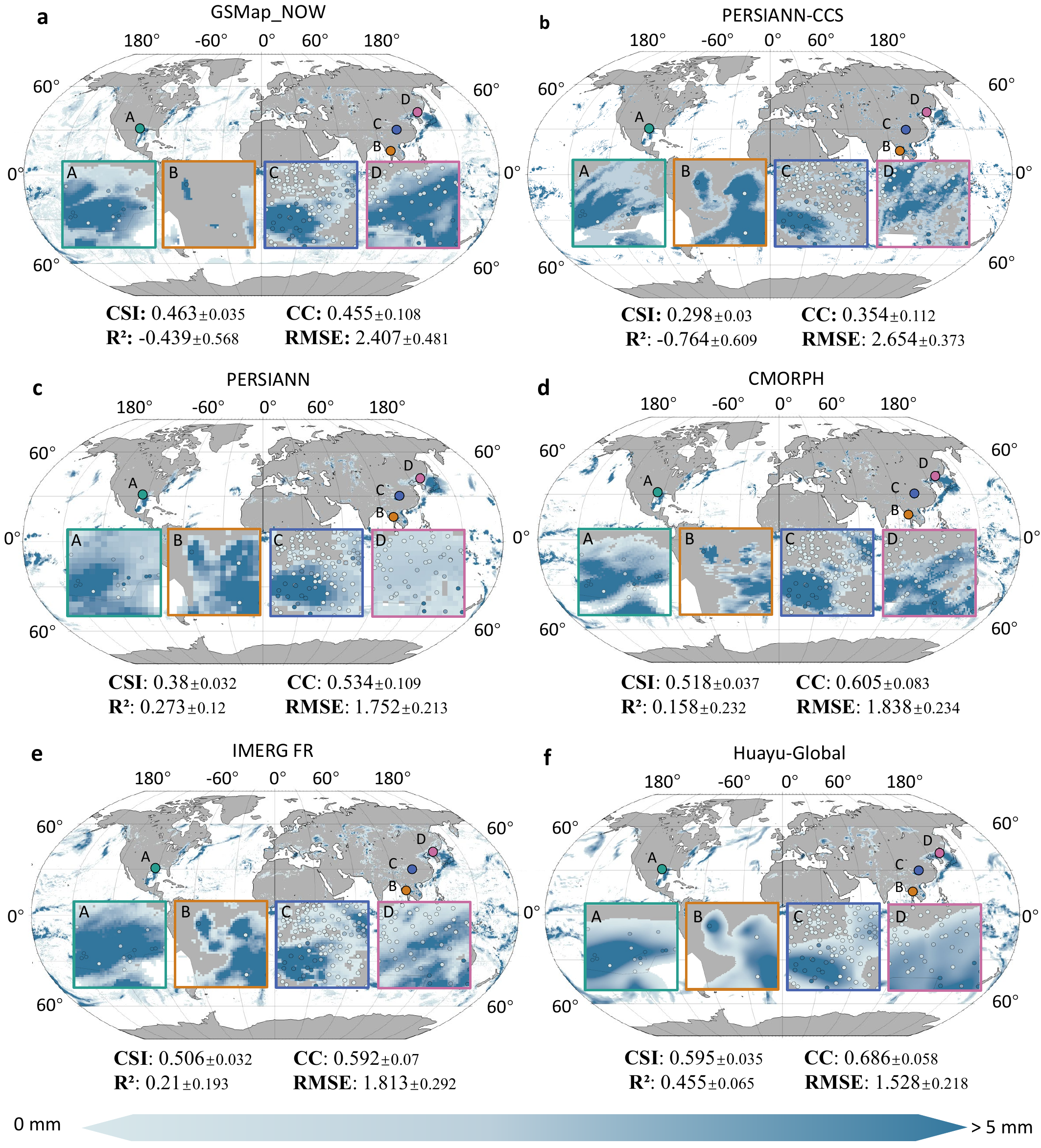}
\caption{Representative spatial comparison of precipitation estimates for one selected evaluation case. Panels show (a) GSMaP\_NOW, (b) PERSIANN-CCS, (c) PERSIANN, (d) CMORPH, (e) IMERG FR, and (f) Huayu-Global. Each panel presents a quasi-global precipitation field with four enlarged regional views; station markers indicate the validation locations in the zoomed regions. The annotations report case-specific CSI, CC, R$^2$, and RMSE for the displayed case only and should not be interpreted as the full-sample scores in Table~\ref{tab:metrics_selected}.}
\label{fig:visual}
\end{figure}

The selected case in Fig.~\ref{fig:visual} illustrates the spatial character of the retrievals. All six products locate the main precipitation systems in broadly similar areas, but Huayu-Global preserves connected rain bands and internal intensity gradients across the mosaic, whereas several baselines produce fragmented, block-like, or isolated features. The four enlarged regions show the same contrast for both elongated rain bands and localized precipitation centres.

The case-specific metrics are consistent with the visual result: Huayu-Global has the highest CSI (0.595), CC (0.686), and R$^2$ (0.455), together with the lowest RMSE (1.528~mm), among the six displayed products. Because this is a single selected scene and some uncertainty ranges overlap, it illustrates rather than establishes the product-wide ranking; that conclusion rests on the full-sample evaluation in Table~\ref{tab:metrics_selected}.

\subsection{Regional performance}

% Requires: \usepackage{booktabs}
% \newcommand{\tabincell}[2]{\begin{tabular}{@{}#1@{}}#2\end{tabular}}
\begin{table*}[htbp!]
  \scriptsize
  \setlength{\extrarowheight}{2pt}
  \centering
  \caption{Regional performance of Huayu-Global across three operational sectors and the quasi-global station sample. Metrics are evaluated over July--December~2022 and January--June~2025. Sector entries are record-count-weighted summaries of independently evaluated longitude segments and are not recomputed from pooled sector samples; the quasi-global entry is computed directly over the full evaluation sample. Because R$^2$ is computed relative to the range of the records being pooled, and the quasi-global R$^2$ is computed over the full range of the pooled records, it is not a record-count-weighted average of sector R$^2$ values; between-sector differences in observed precipitation can therefore make the pooled quasi-global R$^2$ exceed the individual sector values.}
  \label{tab:parts}
  \begin{tabular}{p{2.5cm}<{\centering} p{1.6cm}<{\centering} p{1.2cm}<{\centering} p{1.2cm}<{\centering} p{1.2cm}<{\centering} p{1.5cm}<{\centering}}
    \toprule
    \tabincell{c}{\textbf{Region}}
    &\tabincell{c}{\textbf{Records}}
    &\tabincell{c}{\textbf{CSI}$\uparrow$}
    &\tabincell{c}{\textbf{ACC}$\uparrow$}
    &\tabincell{c}{\textbf{R$^2$}$\uparrow$}
    &\tabincell{c}{\textbf{RMSE}$\downarrow$\\(mm)} \\
    \midrule
    \tabincell{c}{Americas}
    &\tabincell{c}{302,755}
    &\tabincell{c}{0.495}
    &\tabincell{c}{0.814}
    &\tabincell{c}{0.339}
    &\tabincell{c}{2.452} \\
    \tabincell{c}{Europe-Africa}
    &\tabincell{c}{135,130}
    &\tabincell{c}{0.398}
    &\tabincell{c}{0.888}
    &\tabincell{c}{0.247}
    &\tabincell{c}{1.597} \\
    \tabincell{c}{Asia-Pacific}
    &\tabincell{c}{1,410,082}
    &\tabincell{c}{0.415}
    &\tabincell{c}{0.914}
    &\tabincell{c}{0.356}
    &\tabincell{c}{1.866} \\
    \midrule
    \tabincell{c}{Quasi-global}
    &\tabincell{c}{1,847,967}
    &\tabincell{c}{0.441}
    &\tabincell{c}{0.896}
    &\tabincell{c}{0.362}
    &\tabincell{c}{1.961} \\
    \bottomrule
  \end{tabular}
\end{table*}

Regional summaries are spatially heterogeneous and mainly correspond to the three sensor domains. The Americas sector (GOES ABI) has the highest CSI (0.495) and the highest RMSE (2.452~mm); the Europe-Africa sector (Meteosat SEVIRI) has the lowest RMSE (1.597~mm) and the lowest sector-level CSI (0.398); and the Asia-Pacific sector (FY-4B AGRI) has the largest validation sample (1,410,082 records), the highest ACC (0.914), and the highest sector-level R$^2$ (0.356).

\begin{figure}[!htbp]
\centering
\includegraphics[width=\linewidth]{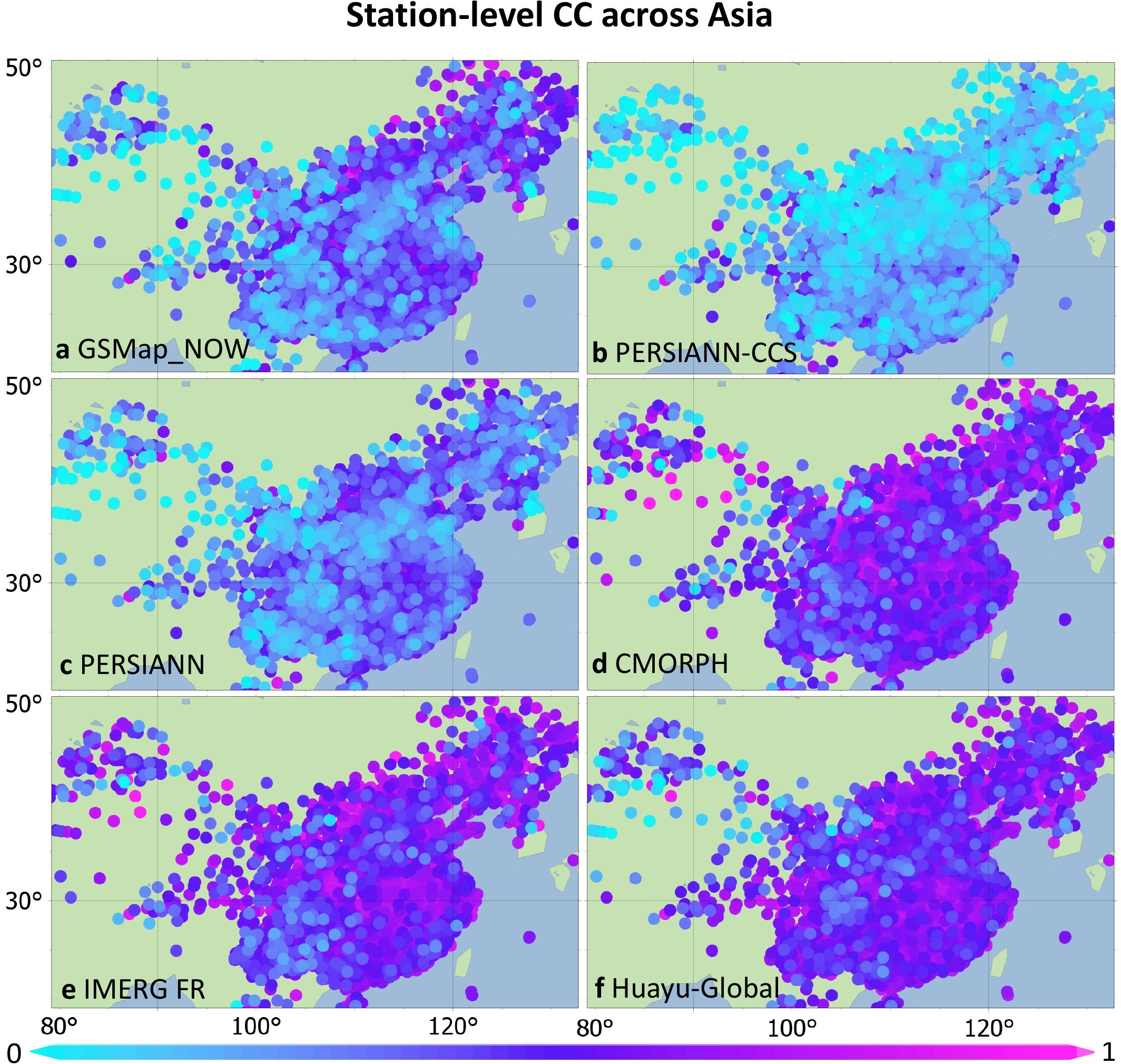}
\caption{Station-level correlation coefficient (CC) across the Asia validation domain, shown for the station network concentrated in East Asia. Panels~a--f show GSMaP\_NOW, PERSIANN-CCS, PERSIANN, CMORPH, IMERG FR, and Huayu-Global, respectively. Each circle marks a validation station and is coloured by the Pearson CC between the product estimates and station observations, calculated from all available paired 3-hour records in the July--December~2022 and January--June~2025 validation windows. The colour scale ranges from 0 (cyan) to 1 (magenta).}
\label{fig:station_cc_asia}
\end{figure}

Figure~\ref{fig:station_cc_asia} provides station-level spatial detail for the Asia validation domain. The corresponding station-level CC maps for the Americas and Europe domains are provided in Extended Data Figs.~\ref{fig:station_cc_americas} and~\ref{fig:station_cc_europe}, respectively. Station-level CSI maps for the Americas, Europe, and Asia domains are provided in Extended Data Figs.~\ref{fig:station_csi_americas}--\ref{fig:station_csi_asia}. A paired comparison across common stations and identical precipitation datasets showed that CC exceeded CSI in 70.54\% of the station--dataset pairs, and the station-averaged CC was higher than the corresponding station-averaged CSI for all six datasets. These absolute values are not directly comparable, however, because CC and CSI have distinct definitions and differ in their sensitivity to zero-precipitation records, missed events, and false alarms. The comparison therefore characterizes the relationship between the two diagnostics rather than indicating better model performance from the metric with the larger numerical value.

\subsection{Performance across validation conditions}

\begin{figure}[!htbp]
\centering
\includegraphics[width=1\textwidth]{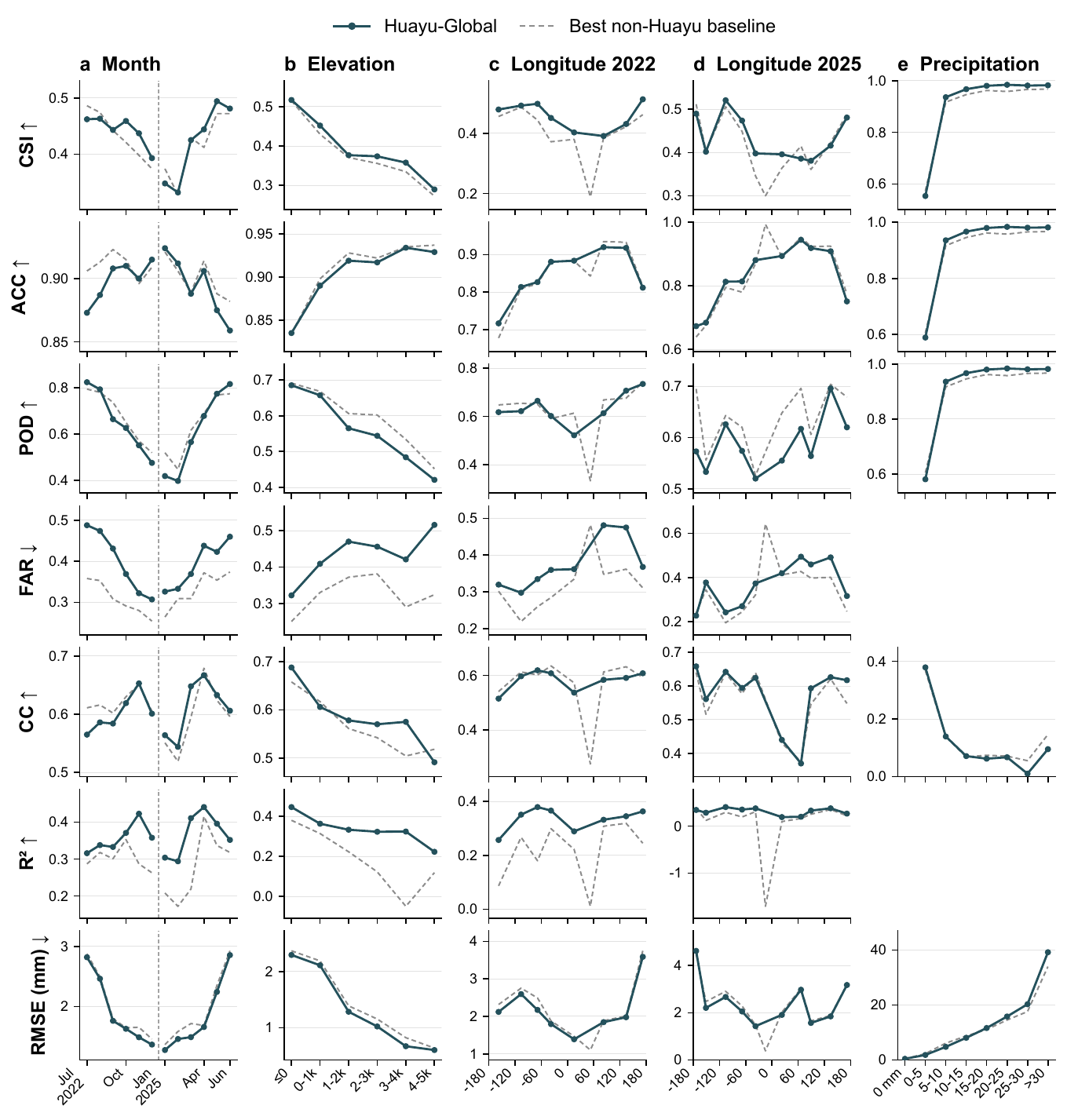}
\caption{Stratified performance across validation conditions. Rows show CSI, ACC, POD, FAR, CC, R$^2$, and RMSE; columns show validation month, station elevation, longitude sector in 2022 and 2025, and observed 3-hour precipitation amount. Solid teal lines denote Huayu-Global, and dashed grey lines denote the strongest non-Huayu baseline selected separately within each bin and metric. Metric--bin combinations that are undefined or uninformative by construction are omitted.}
\label{fig:combined_metrics}
\end{figure}

Figure~\ref{fig:combined_metrics} resolves the aggregate comparison across temporal, geographic, and precipitation regimes. Across validation months, Huayu-Global generally remains close to or exceeds the best-baseline envelope, although the size of the difference varies by metric and month. The 2022 longitude profiles are comparatively stable, whereas the 2025 profiles are more heterogeneous; Huayu-Global nevertheless remains among the leading products across most valid longitude sectors.

Stratification also identifies the most challenging evaluation regimes. CSI and POD generally decrease with station elevation while FAR increases, indicating weaker event detection over high terrain. RMSE rises sharply with observed precipitation amount, identifying heavy precipitation as a prominent regime of increasing magnitude error. Because the strongest baseline is selected separately within each bin and metric, the grey series is an upper-envelope diagnostic rather than a comparison with one fixed product; amount-conditioned categorical metrics likewise describe the selected precipitation bins rather than an independent threshold-sensitivity test.

\clearpage

\section{Discussion}

Huayu-Global extends the earlier regional Huayu framework~\citep{song2026huayu} from a single-sensor regional retrieval to a three-sensor quasi-global system covering approximately 85\% of Earth's land surface between 60$^\circ$S and 60$^\circ$N. It demonstrates that high-frequency geostationary imagery can deliver real-time-input precipitation estimates with skill previously associated with delayed post-processed products. At independent station locations, Huayu-Global achieves the highest CSI and R$^2$ and the lowest RMSE among the evaluated products. Relative to GSMaP\_NOW, it increases CSI by 34.9\% and reduces RMSE by 31.8\%. The central advance is the conversion of continuously available cloud-evolution observations into high-skill precipitation information within the operational decision window.

This capability connects satellite retrieval directly to time-critical environmental workflows. Huayu-Global generates a new 0.05$^\circ$ precipitation field every 30~minutes, and data loading and inference require less than 5~minutes per timestep after model initialization on a single workstation (Methods). The resulting fields can provide a timely observation layer for flood nowcasting, hydrological forecasting, agricultural monitoring and disaster response. They also complement recent deep-learning nowcasting and weather-prediction systems, which require rapidly updated precipitation information to initialize, constrain and evaluate forecasts~\citep{sonderby2020metnet,ravuri2021skilful,espeholt2022deep,zhang2023skilful}.

The cross-constellation design makes this capability scalable. The three sector models share the same Head--Geographic Attention Block (GeoAB)--Tail architecture while using separate parameters for ABI, SEVIRI, and AGRI streams. This combination applies a consistent representation pathway to cloud evolution while adapting to differences in spectral coverage, resolution, scan frequency and viewing geometry. All three models achieve useful retrieval skill within their respective domains (Table~\ref{tab:parts}), enabling their outputs to form a coherent quasi-global product. The same design provides a practical path for incorporating Meteosat Third Generation, FY-4C, and Geostationary Extended Observations (GeoXO) missions as the global geostationary constellation evolves.

Training on 2023--2024 data and evaluation in July--December~2022 and January--June~2025 show that the retrieval skill extends beyond the training years. The aggregate advantage persists across these windows, while analyses by month, station elevation, longitude sector and precipitation amount (Fig.~\ref{fig:combined_metrics}) identify where that skill varies. The Americas sector (GOES ABI) achieves both the highest CSI (0.495) and highest RMSE (2.452~mm), consistent with strong event detection in a domain containing intense convective rainfall. The Europe-Africa sector (Meteosat SEVIRI) has the lowest CSI (0.398) and lowest RMSE (1.597~mm), indicating more conservative detection and smaller magnitude errors. These regional contrasts connect retrieval performance to sensor characteristics, climate regimes and the uneven distribution of station samples.

Huayu-Global is supervised with IMERG FR yet exceeds it for CSI, R$^2$, and RMSE at independent station locations. This result shows that a gridded supervision product need not define the performance ceiling of a retrieval evaluated at a different spatial scale. The 0.05$^\circ$ grid, explicit separation of rain occurrence and amount~\citep{song2026huayu}, and high-frequency cloud-evolution sequences allow Huayu-Global to sharpen rainfall boundaries, timing and local magnitude relative to the 0.1$^\circ$ training target. Independent radar composites, additional dense gauge networks and oceanic observations would extend this evidence to spatial regimes that remain weakly sampled by the present evaluation.

The remaining limitations define the next scientific targets. Geostationary coverage is restricted to approximately 60$^\circ$S--60$^\circ$N, leaving high latitudes dependent on polar-orbiting observations. The Hadley Centre Integrated Surface Database (HadISD)~\citep{dunn2012hadisd,smith2011integrated,dunn2014pairwise,dunn2016expanding} and China Meteorological Administration (CMA) networks provide extensive land validation but sparse coverage over Africa, South America and the oceans. Warm rain, snow, orographic precipitation and intense convection also remain challenging because cloud-top radiances do not uniquely determine surface rainfall~\citep{wang2021infrared,ma2022fy4qpe,zhu2024pecam,yang2026snow}. These gaps motivate complementary observations and retrieval constraints rather than further expansion of the same satellite-only formulation.

The diagnostic patterns provide a roadmap for that expansion. Elevation-dependent degradation (Fig.~\ref{fig:combined_metrics}, column~b) makes terrain, wind fields and orographic constraints the first priority, while intensity-dependent RMSE growth (Fig.~\ref{fig:combined_metrics}, column~e) motivates loss reweighting and passive-microwave fusion for heavy rainfall. Ensemble uncertainty estimates and robustness to missing scans would make the fields more useful for risk-sensitive decisions, and polar-orbiting microwave observations could extend retrieval towards higher latitudes. Combined with the rapid progress of data-driven global weather prediction~\citep{bi2023accurate,chen2023fuxi}, these developments could turn cross-constellation precipitation retrieval into a continuously updated observation layer linking satellite systems, hydrological models and early-warning services. Huayu-Global provides the present foundation for that transition from delayed precipitation archives to active Earth-system prediction.

\section{Methods}

\subsection{Preprocessing}
The preprocessing stage summarized in Fig.~\ref{fig:processing} converts satellite observations from different geostationary platforms into georeferenced multispectral cloud-feature tensors and reprojects them to a common 0.05$^\circ$ grid. Channel-wise normalization and feature standardization are applied separately for ABI, SEVIRI, and AGRI to reduce cross-sensor heterogeneity while preserving sensor-specific information. Invalid pixels and low-quality retrieval regions are filled by spatial interpolation from valid neighbouring pixels during preprocessing; regions with persistent data gaps spanning the full interpolation window are flagged as missing and excluded from both training and evaluation. Training pairs are generated by collocating geostationary feature tensors with IMERG FR precipitation fields at matching timestamps; at each 30-min interval, one IMERG FR field is paired with two AGRI or SEVIRI scans or three ABI scans according to the native acquisition frequency. At inference time, sector outputs are transformed to the common grid and mosaicked with overlap-aware blending, in which adjacent sector outputs are combined by averaging across overlap zones at mosaic boundaries.

\subsection{Model framework}
Huayu-Global maps multispectral cloud observations to 30-min precipitation estimates through three independently trained sector models. The models share the same Head--GeoAB--Tail architecture, where GeoAB denotes Geographic Attention Block (Extended Data Fig.~\ref{fig:network})~\citep{song2025mdg625,song2026geoattx,song2026huayu}, but have separate parameters for ABI, SEVIRI, and AGRI streams. This design keeps the retrieval framework consistent while allowing each model to adapt to its sensor-domain distribution. The choice of a shared architecture with independent weights reflects the hypothesis that cloud-to-rainfall relationships share common physical structure across sensors---captured by the common Head--GeoAB--Tail design---but require sensor-specific parameterization of spectral response functions, viewing geometry, and scan cadence---captured by separately trained weights. The GeoAB modules combine attention mechanisms introduced by the Transformer~\citep{vaswani2017attention}, shifted-window attention derived from Swin Transformer and SwinIR designs~\citep{liu2021swin,liu2022swin,liang2021swinir}, and long-range attention components developed in earlier geographic image-restoration and precipitation-retrieval work~\citep{song2022direction,song2022lightweight,song2025mdg625,song2026huayu}. A single unified model would need to internally reconcile conflicting sensor characteristics, while three entirely separate architectures would forgo the modelling efficiency and cross-sensor consistency that a common framework provides. The sector-wise outputs are merged into a spatially continuous quasi-global product after reprojection and overlap blending. Six consecutive 30-min estimates are accumulated for evaluation against 3-hour station records.

\subsection{Runtime performance}

Single-machine runtime was measured on a 64-bit Windows~11 workstation equipped with an Intel Core i7-10700F CPU (8 cores and 16 threads), 48~GB of system memory, and one NVIDIA GeForce RTX~4060~Ti GPU with 8~GB of memory. For a representative single-timestamp Huayu-Global run with the required satellite files stored locally, model initialization and weight loading took 91.5~s, input-data loading took 195.0~s, and retrieval inference took 95.2~s. The complete first run therefore required 381.7~s (6.36~min). Because model loading is a one-time startup cost when the process remains active, the measured data-loading and inference stages for a subsequent timestamp total 290.2~s (4.84~min). These measurements characterize local computational performance and exclude upstream satellite-data delivery, local transfer, quality control, and downstream product dissemination. In this manuscript, ``real-time'' therefore means that inference uses only observations available for the target time and that local processing is short relative to the 30-min product cycle; it does not claim a measured end-to-end public-release latency. Faster storage or multi-GPU deployment could further reduce data loading time or support concurrent processing.

\subsection{Training and optimization}

The training data cover 2023 and 2024, whereas the independent station validation windows are July--December~2022 and January--June~2025. This temporal separation is used to evaluate generalization outside the training years. Huayu-Global adopts the dual-target loss introduced in the earlier Huayu framework~\citep{song2026huayu}. The objective addresses the imbalance between non-raining and raining pixels by separating precipitation occurrence from precipitation amount. For each training sample $i$, let $r_i$ be the IMERG FR precipitation target, $\hat{r}_i$ the regression output, $z_i=\mathbb{F}(r_i\geq0.1~\mathrm{mm})$ the rain-occurrence label, and $\hat{\mathbf{q}}_i$ the predicted rain--no-rain probability vector. The training loss is

\begin{equation}
\mathcal{L}
=\frac{1}{\max(N_+,1)}\sum_{i=1}^{N} z_i\,\operatorname{SmoothL1}(\hat{r}_i,r_i)
+\frac{1}{N}\sum_{i=1}^{N}\operatorname{CE}(\hat{\mathbf{q}}_i,z_i),
\end{equation}

where $N$ is the number of pixels in a training batch, $N_+=\sum_i z_i$, $\operatorname{CE}$ denotes cross-entropy loss, and the SmoothL1 term is evaluated only over positive-rainfall pixels. At inference, pixels classified as non-raining are set to zero, while pixels classified as raining retain the regression estimate. Space-time samples are shuffled during training, and model checkpoints are selected according to held-out validation performance. Because IMERG FR is the supervision target and is also one of the delayed comparison products, the product-ranking claims are solely based on independent station observations rather than on agreement with the training target.

All three sector models use AdamW optimization~\citep{loshchilov2017decoupled}, a batch size of 16, and 100 training epochs on four NVIDIA RTX 6000 Ada GPUs with 48 GB memory each. The implementation uses PyTorch with Python 3.12 on Linux. The three models share the same architecture and parameter count, while the learning rate and decay schedule are tuned separately for ABI, AGRI, and SEVIRI streams (Extended Data Table~\ref{tab:config}).

\subsection{Evaluation metrics}

Evaluation is performed on matched station-product records at 3-hour accumulation scale, with each record constructed from six consecutive 30-min precipitation fields. For 3-hour evaluation, hourly and sub-hourly station records are aggregated to the corresponding 3-hour windows. All products are evaluated over the same valid station-time set after applying common quality-control and missing-data filters, so differences in record count do not drive the product rankings in Table~\ref{tab:metrics_selected}. Product values are extracted from each product's native grid at the station coordinates before temporal accumulation, avoiding an artificial common-grid resampling step that could smooth or sharpen products differently. For 30-min products, six consecutive fields are summed to produce the 3-hour accumulation; for products natively delivered at coarser temporal resolution, the accumulation is computed over the number of constituent fields that fall within the 3-hour window. The two validation periods comprise 730 candidate windows: 00:00--03:00 and 12:00--15:00~UTC on each day from 1 July to 31 December 2022 and from 1 January to 30 June 2025. Candidate windows for which missing source satellite imagery prevented consistent generation of the corresponding gridded precipitation fields were excluded before station matching. For the remaining windows, a station-time sample was retained only when a quality-controlled gauge accumulation was available and could be collocated with all seven comparison products (PERSIANN-CCS, GSMaP\_NOW, CMORPH, PERSIANN, and IMERG ER, LR, and FR). Missing, non-finite, or product-defined invalid precipitation values were excluded from metric calculations, whereas valid zero-precipitation values were retained. Consequently, the record count for an individual station represents the number of valid gauge--grid collocations rather than the 730 theoretical windows; among the 5,395 stations, the largest station-level count was 569. Precipitation occurrence is evaluated at the native subperiods used to build each Huayu-Global 3-hour record using a 0.1~mm threshold before contingency counts are accumulated over the validation sample; the 3-hour intensity threshold is also set to 0.1~mm for all products. Detection skill is assessed by critical success index (CSI), probability of detection (POD), false-alarm ratio (FAR), and accuracy (ACC). Let $\mathrm{TP}$, $\mathrm{FN}$, $\mathrm{FP}$, and $\mathrm{TN}$ denote true positives, false negatives, false positives, and true negatives, respectively. The categorical metrics are computed as

\begin{equation}
\begin{aligned}
\mathrm{CSI}&=\frac{\mathrm{TP}}{\mathrm{TP}+\mathrm{FN}+\mathrm{FP}}, & \mathrm{POD}&=\frac{\mathrm{TP}}{\mathrm{TP}+\mathrm{FN}},\\
\mathrm{FAR}&=\frac{\mathrm{FP}}{\mathrm{TP}+\mathrm{FP}}, & \mathrm{ACC}&=\frac{\mathrm{TP}+\mathrm{TN}}{\mathrm{TP}+\mathrm{FN}+\mathrm{FP}+\mathrm{TN}}.
\end{aligned}
\end{equation}

Quantitative consistency is assessed by CC, R$^2$, and RMSE. For product estimates $p_i$ and station observations $o_i$, both expressed as 3-hour accumulated precipitation amounts,

\begin{equation}
\begin{aligned}
\mathrm{RMSE}&=\sqrt{\frac{1}{n}\sum_{i=1}^{n}(p_i-o_i)^2},\\
R^2&=1-\frac{\sum_{i=1}^{n}(p_i-o_i)^2}{\sum_{i=1}^{n}(o_i-\bar{o})^2},\\
\mathrm{CC}&=\frac{\sum_{i=1}^{n}(p_i-\bar{p})(o_i-\bar{o})}{\sqrt{\sum_{i=1}^{n}(p_i-\bar{p})^2}\sqrt{\sum_{i=1}^{n}(o_i-\bar{o})^2}},
\end{aligned}
\end{equation}
where $\bar{p}$ and $\bar{o}$ are the mean product estimate and mean station observation.

\subsection{Statistics and uncertainty}
Uncertainty ranges reported with the evaluation metrics are computed from 1,000 bootstrap resamples. A resampling block is defined by each station and UTC period and contains all valid 3-hour records. For each bootstrap iteration, station-time blocks are drawn with replacement and metrics are recomputed on the resampled set. The bootstrap distribution is summarized by the 2.5th and 97.5th percentiles, giving a central 95\% confidence interval. Compact tables report this uncertainty as $\pm$ one-half of the central 95\% interval. These intervals quantify sampling uncertainty over the matched station network and do not include structural uncertainty from product algorithms, gauge undercatch, representativeness error between point gauges and grid cells, or unobserved oceanic regions.

\subsection{Data}
Huayu-Global uses three operational geostationary satellite systems to cover most low- and mid-latitude regions (60$^\circ$S--60$^\circ$N): GOES-16/18/19 Advanced Baseline Imager (ABI) for the Americas sector~\citep{schmit2005introducing,schmit2017closer}, Meteosat-9/10/11 Spinning Enhanced Visible and Infrared Imager (SEVIRI) for the Europe-Africa sector~\citep{schmetz2002introduction}, and FY-4B Advanced Geostationary Radiation Imager (AGRI) for the Asia-Pacific sector~\citep{yang2017introducing}. GOES-16 and GOES-19 are treated as the same ABI stream, and Meteosat-10 and Meteosat-11 are treated as the same SEVIRI stream, with operational transitions handled through provider data records and sector footprint metadata. All source observations are transformed onto a common 0.05$^\circ$ analysis grid and harmonized in time before model inference. Detailed satellite specifications, data volumes, and provider information are given in Extended Data Tables~\ref{tab:satellite} and~\ref{tab:datasize}.

Huayu-Global is benchmarked against seven representative satellite precipitation products spanning infrared-based, passive-microwave-informed, morphing-based, multi-sensor, real-time, and delayed post-processed retrieval families: GSMaP\_NOW~\citep{kubota2020global}, PERSIANN-CCS and PERSIANN~\citep{nguyen2018persiann}, CMORPH~\citep{joyce2004cmorph}, and IMERG ER/LR/FR~\citep{huffman2015nasa,tan2019imerg,huffman2023imerg}. This broad selection provides a comparison set for evaluating both event detection and quantitative precipitation estimation.

Independent in situ observations from the HadISD global station dataset~\citep{dunn2012hadisd,smith2011integrated,dunn2014pairwise,dunn2016expanding} and CMA station network provide the validation data, with 5,395 stations and 1,847,967 three-hourly records. IMERG FR fields are used as gridded training targets, whereas ground observations are reserved exclusively for independent validation and cross-product comparison. This separation reduces circularity between model supervision and final performance assessment, but the station network is unevenly distributed and does not directly evaluate open-ocean precipitation.


\section*{Data availability}

The data products used in this study are listed in Extended Data Table~\ref{tab:datasize}. GOES-R ABI Level-1b observations are publicly available from the National Oceanic and Atmospheric Administration National Centers for Environmental Information (NOAA NCEI; \url{https://www.ncei.noaa.gov/products/goes-terrestrial-weather-abi-glm}); Meteosat SEVIRI Level-1.5 observations are available through the European Organisation for the Exploitation of Meteorological Satellites (EUMETSAT) Data Store (\url{https://data.eumetsat.int/}); and FY-4B AGRI Level-1 observations are available from the National Satellite Meteorological Center (NSMC; \url{https://www.nsmc.org.cn/nsmc/en/home/index.html}). Provider registration and acceptance of the applicable data-use terms may be required.

The comparison precipitation products can be obtained from their original archives: GSMaP from JAXA Global Rainfall Watch (\url{https://sharaku.eorc.jaxa.jp/GSMaP/}), PERSIANN and PERSIANN-CCS from the Center for Hydrometeorology and Remote Sensing (CHRS) Data Portal (\url{https://chrsdata.eng.uci.edu/}), CMORPH from NOAA NCEI (\url{https://www.ncei.noaa.gov/products/climate-data-records/precipitation-cmorph}), and IMERG from the National Aeronautics and Space Administration (NASA) Goddard Earth Sciences Data and Information Services Center (GES DISC; \url{https://disc.gsfc.nasa.gov/datasets?keywords=IMERG}). HadISD station observations are publicly available from the Met Office Hadley Centre (\url{https://www.metoffice.gov.uk/hadobs/hadisd/}). CMA station observations are subject to provider access restrictions and cannot be redistributed by the authors. The trained model files, normalization data, coverage masks, and demonstration satellite observations for 00:00 UTC on January~2, 2025 are openly available from the download location documented in the Huayu-Global repository (\url{https://github.com/songzijiang/Huayu-Global}). Processed training tensors and matched station-validation records are not publicly redistributed because of their volume and the licensing conditions of the source datasets.

\section*{Code availability}

The Huayu-Global inference and multisatellite mosaicking source code is publicly available at \url{https://github.com/songzijiang/Huayu-Global} under the Apache License 2.0. A permanent archived release for the version corresponding to this manuscript will be deposited on Zenodo soon. The repository provides setup instructions, inference scripts, mosaicking scripts, and links to the trained model bundle and demonstration input data, enabling reproduction of the public inference workflow. Training code paths, restricted CMA station data, and the complete station-validation workflow are not redistributed publicly.

\section*{Acknowledgements}
We thank the providers of the datasets used in this work, including FengYun, GOES, and Meteosat satellite imagery; HadISD and CMA gauge records; and the PERSIANN, CMORPH, GSMaP, and IMERG precipitation products.

\section*{Funding}
This work was supported by the National Natural Science Foundation of China (Grant Nos. W2412140 and 42230505) and the International Research Center of Big Data for Sustainable Development Goals (Grant No. CBAS2022GSP07).

\section*{Author contributions}
Z.S. and M.L. conceived the study. Z.S. wrote the original manuscript. Z.S. and X.S. developed the model methodology. S.Y., Y.L., and T.L. conducted experiments, analyzed the results, and generated the figures and tables. Z.S. and Z.Z. developed the Huayu-Global system. Z.Z. and W.X. contributed to satellite, precipitation-product, and station-data processing. L.Y. and M.L. supervised the project. M.L. acquired funding and revised the manuscript. All authors discussed the results and reviewed and approved the final manuscript.

\section*{Competing interests}
The authors declare no competing interests.

\section*{Correspondence and requests for materials}
Correspondence and requests for materials should be addressed to Min Liu (\url{mliu@geo.ecnu.edu.cn}) or Lina Yuan (\url{lnyuan@geoai.ecnu.edu.cn}).

\clearpage
\section*{Extended Data}
\setcounter{figure}{0}
\setcounter{table}{0}
\renewcommand{\figurename}{Extended Data Fig.}
\renewcommand{\tablename}{Extended Data Table}

\begin{figure}[!htbp]
\centering
\includegraphics[width=0.9\textwidth]{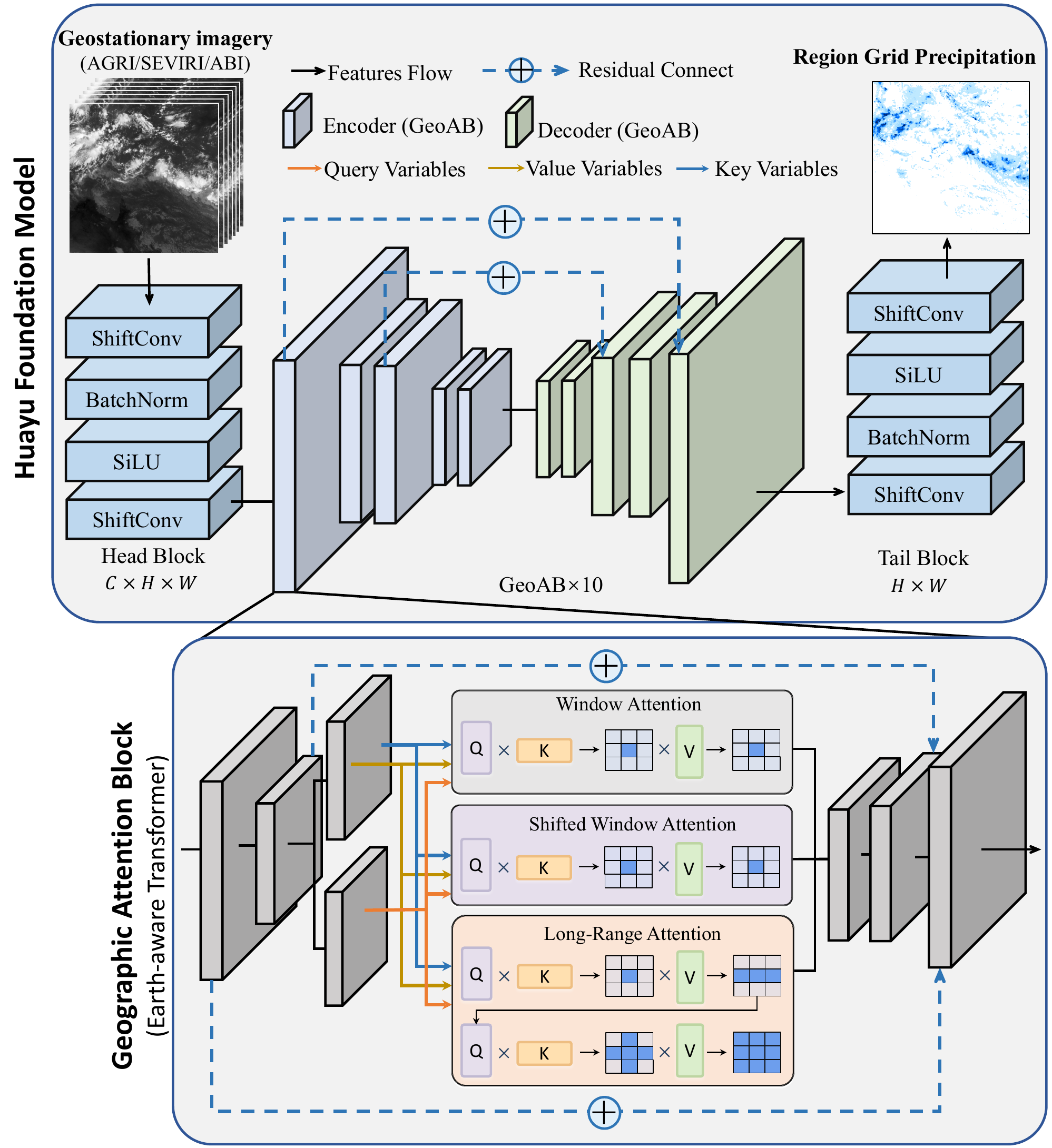}
\caption{Architecture of the Huayu-Global retrieval model. Multispectral geostationary imagery from the AGRI aboard FY-4B, the SEVIRI aboard Meteosat, and the ABI aboard GOES is mapped to regional grid precipitation estimates for the Asia-Pacific, Europe-Africa, and Americas domains. Each sector model contains a Head Block, ten Geographic Attention (GeoAB) modules, and a Tail Block, with residual connections supporting feature propagation. The GeoAB module combines local cloud-texture attention with broader geographic context to capture fine-scale convective signatures and large-scale precipitation patterns.}
\label{fig:network}
\end{figure}

\begin{figure}[!htbp]
\centering
\includegraphics[width=\linewidth]{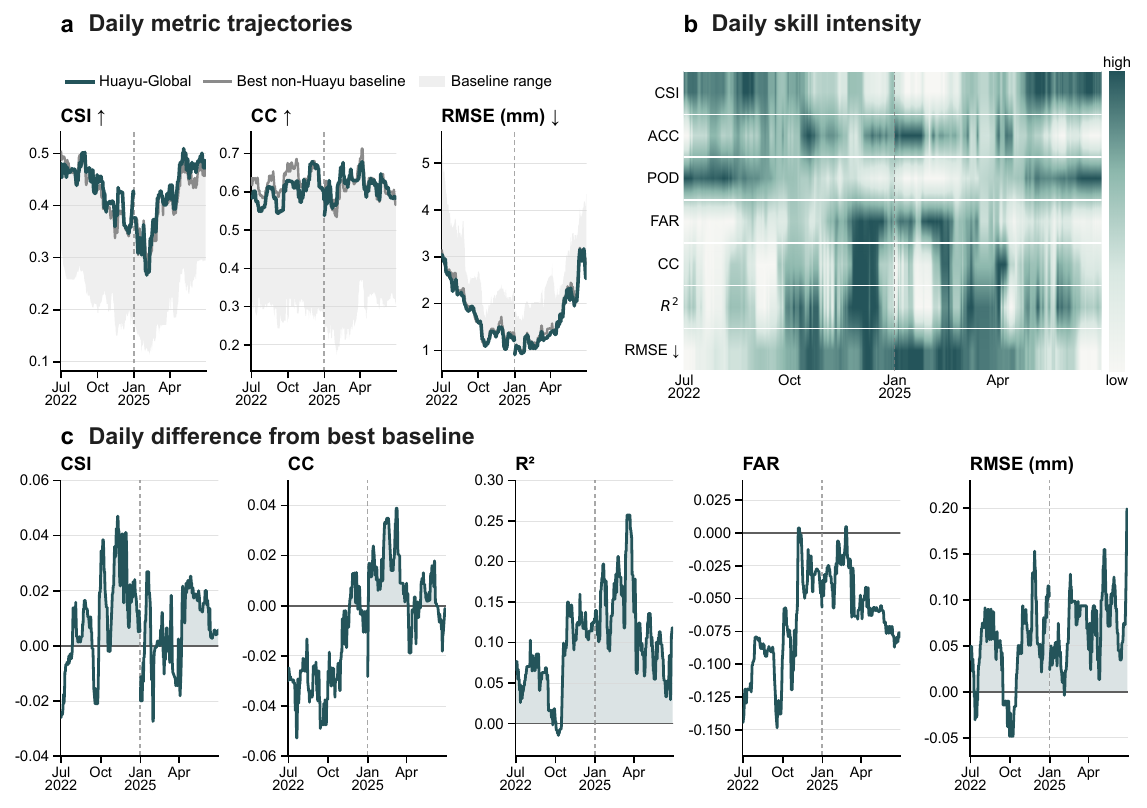}
\caption{Daily-scale stability of Huayu-Global relative to seven baseline precipitation products. Panel~a compares daily CSI, CC, and RMSE trajectories for Huayu-Global, the strongest non-Huayu baseline, and the full baseline range during the July--December~2022 and January--June~2025 validation windows. Panel~b summarizes Huayu-Global's relative daily skill intensity across categorical and continuous metrics. Panel~c shows centred 14-day rolling-median differences between Huayu-Global and the strongest non-Huayu baseline, selected separately for each day and metric. Positive vertical displacement denotes stronger Huayu-Global performance, and teal shading marks positive differences. Dashed vertical lines separate the validation periods. Because the strongest baseline is selected separately by metric and date, this figure is a diagnostic envelope comparison rather than a head-to-head comparison with one fixed product. Huayu-Global generally remains near the upper envelope for CSI and CC and near the lower-error envelope for RMSE, although the strongest non-Huayu baseline changes by metric and date.}
\label{fig:figure_daily_combined}
\end{figure}

\begin{figure}[!htbp]
\centering
\includegraphics[width=\linewidth]{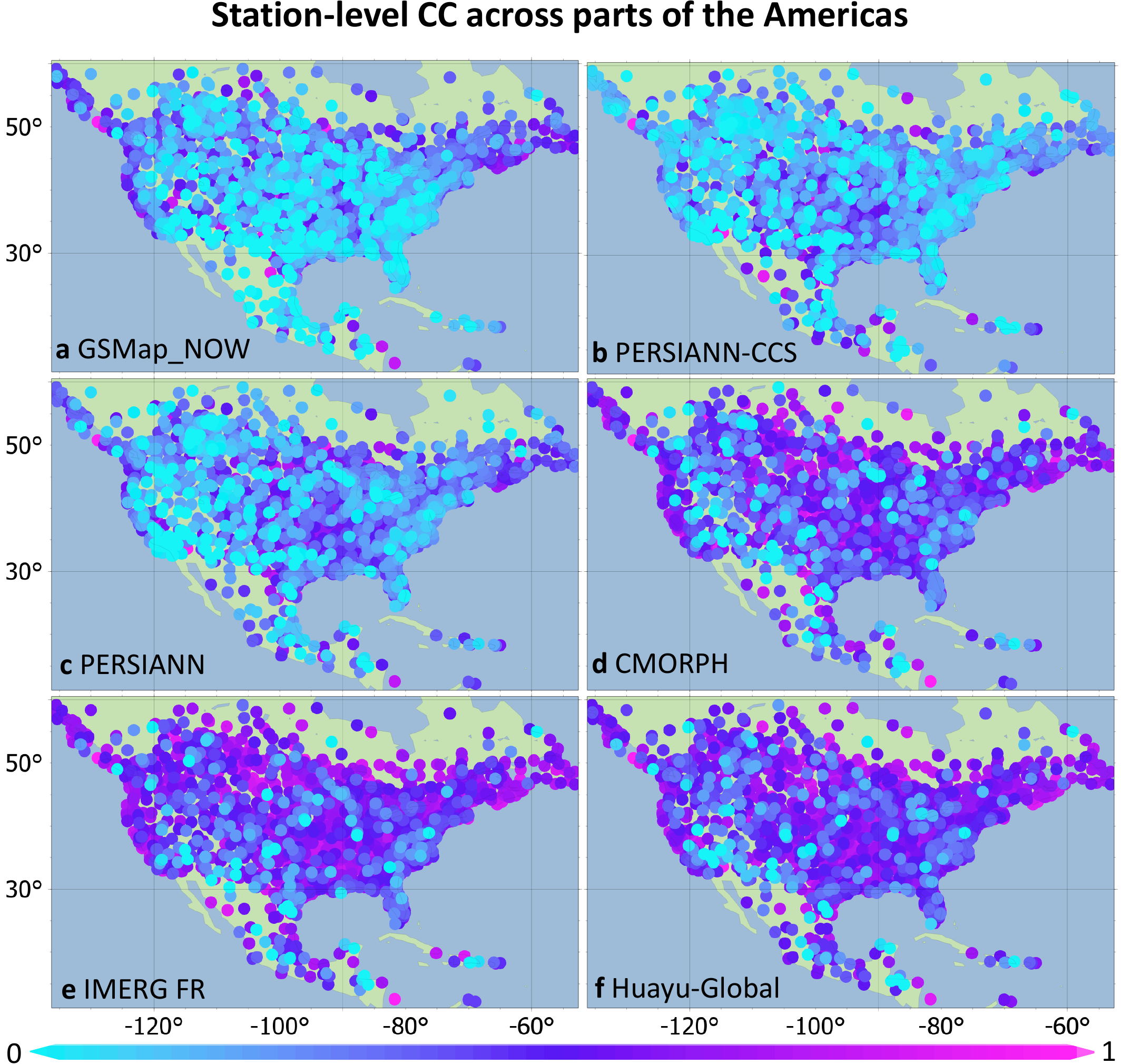}
\caption{The same as Fig.~\ref{fig:station_cc_asia} but across the Americas.}
\label{fig:station_cc_americas}
\end{figure}

\begin{figure}[!htbp]
\centering
\includegraphics[width=\linewidth]{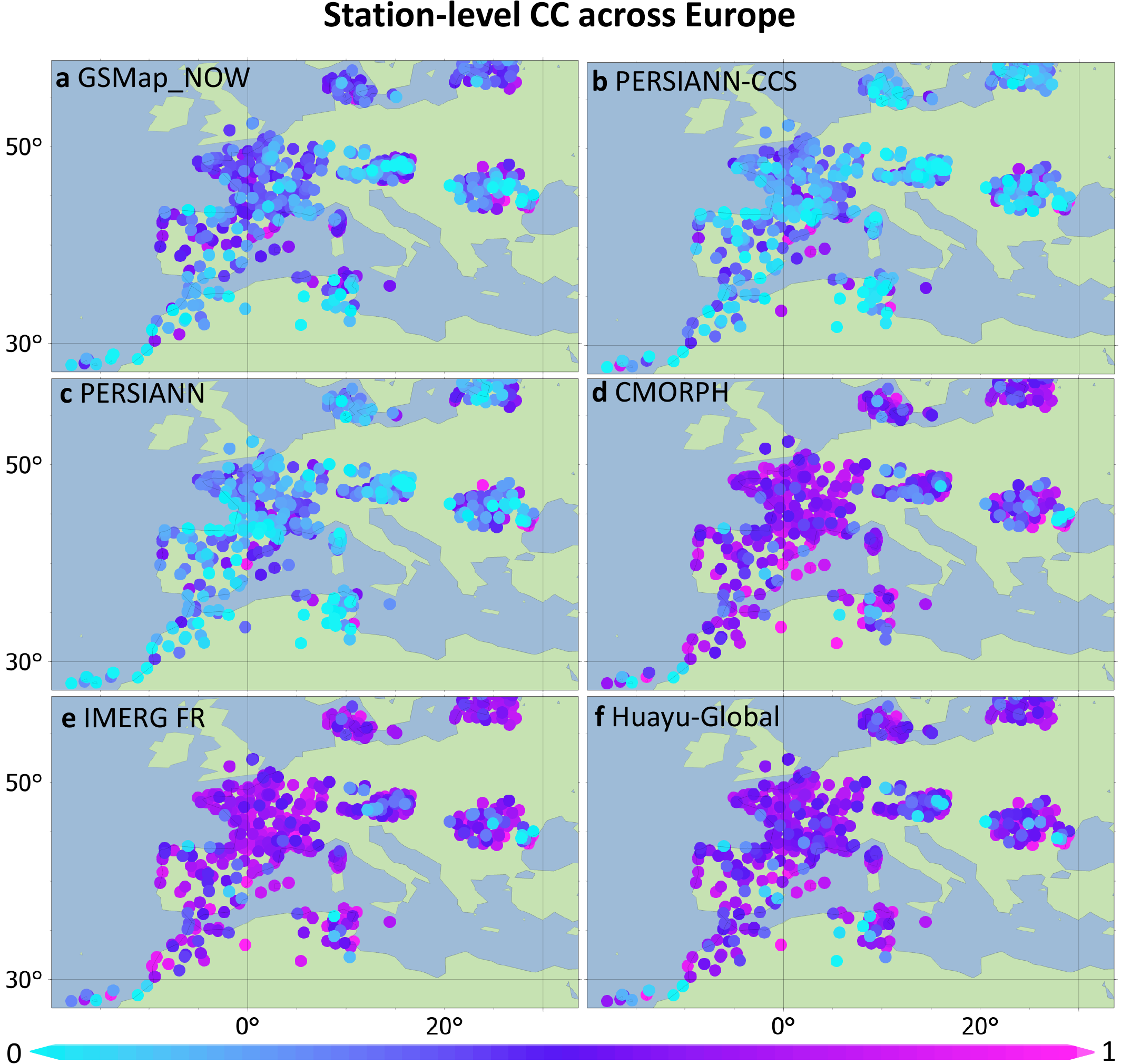}
\caption{The same as Fig.~\ref{fig:station_cc_asia} but across Europe.}
\label{fig:station_cc_europe}
\end{figure}

\begin{figure}[!htbp]
\centering
\includegraphics[width=\linewidth]{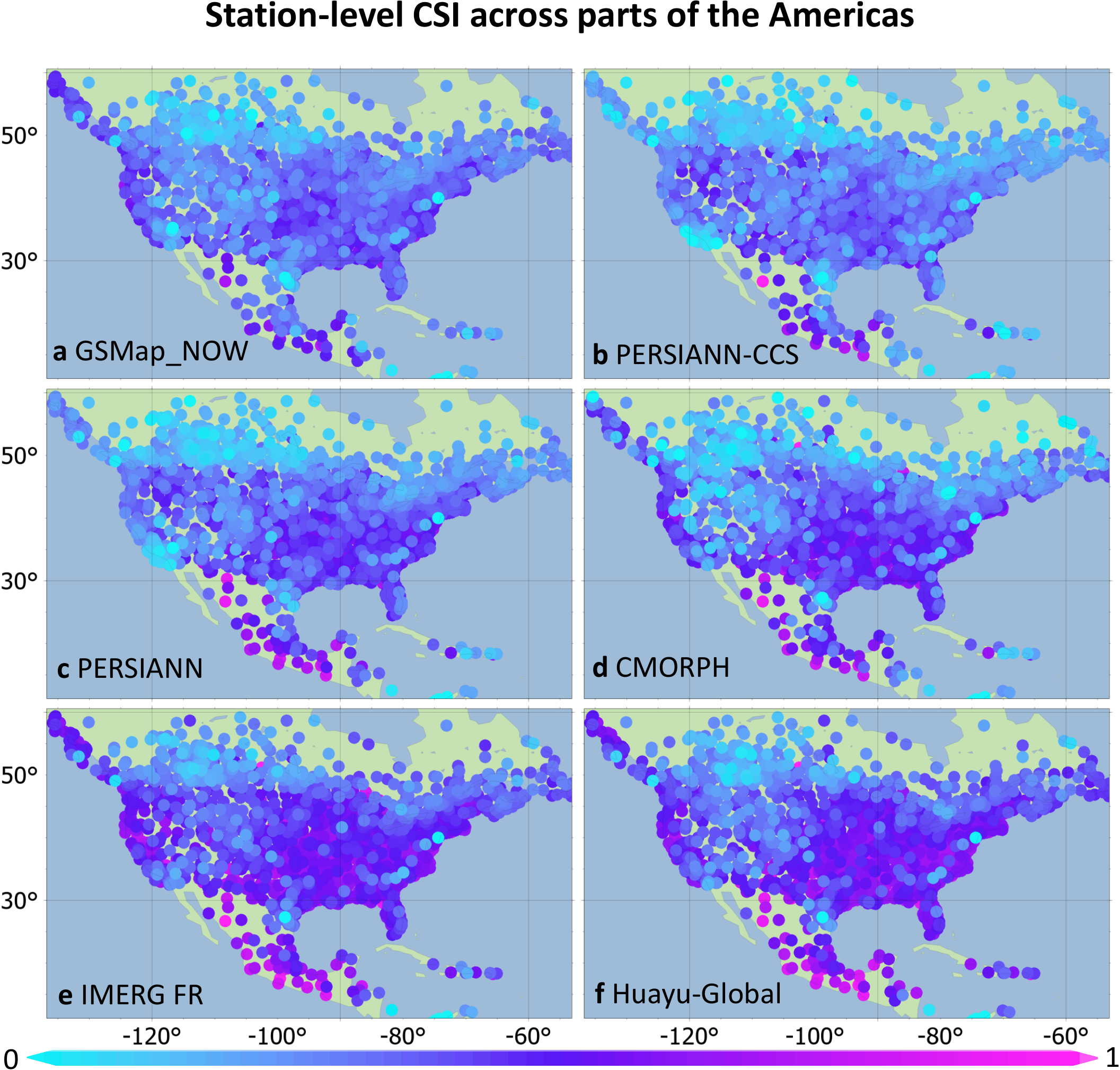}
\caption{Station-level critical success index (CSI) across the Americas validation domain. Panels~a--f show GSMaP\_NOW, PERSIANN-CCS, PERSIANN, CMORPH, IMERG FR, and Huayu-Global, respectively. Each circle marks a validation station and is coloured by the CSI for precipitation-event detection, calculated from all available paired 3-hour records in the July--December~2022 and January--June~2025 validation windows using the common precipitation-occurrence definition specified in Methods. The colour scale ranges from 0 (cyan) to 1 (magenta).}
\label{fig:station_csi_americas}
\end{figure}

\begin{figure}[!htbp]
\centering
\includegraphics[width=\linewidth]{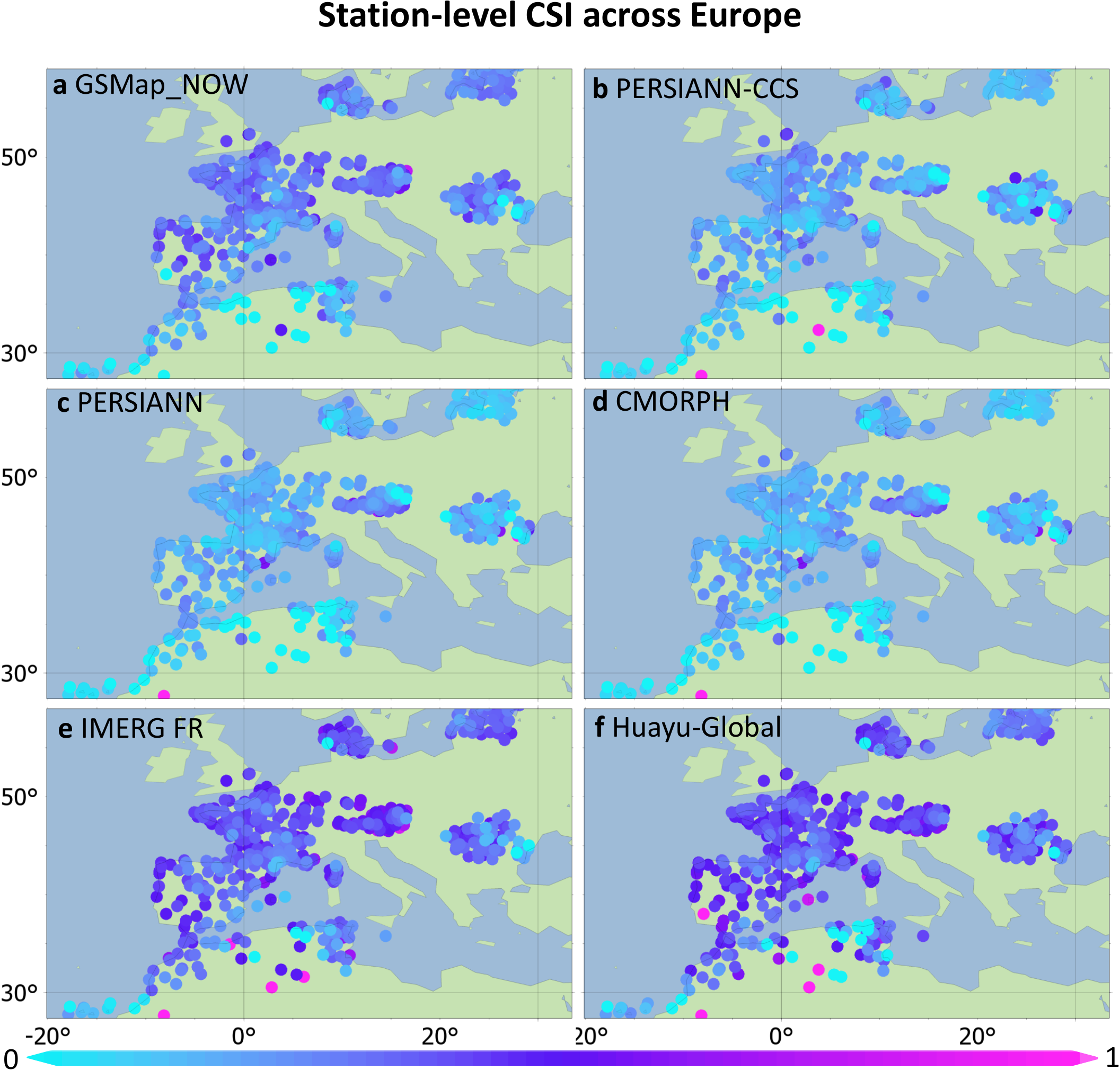}
\caption{The same as Extended Data Fig.~\ref{fig:station_csi_americas} but across Europe.}
\label{fig:station_csi_europe}
\end{figure}

\begin{figure}[!htbp]
\centering
\includegraphics[width=\linewidth]{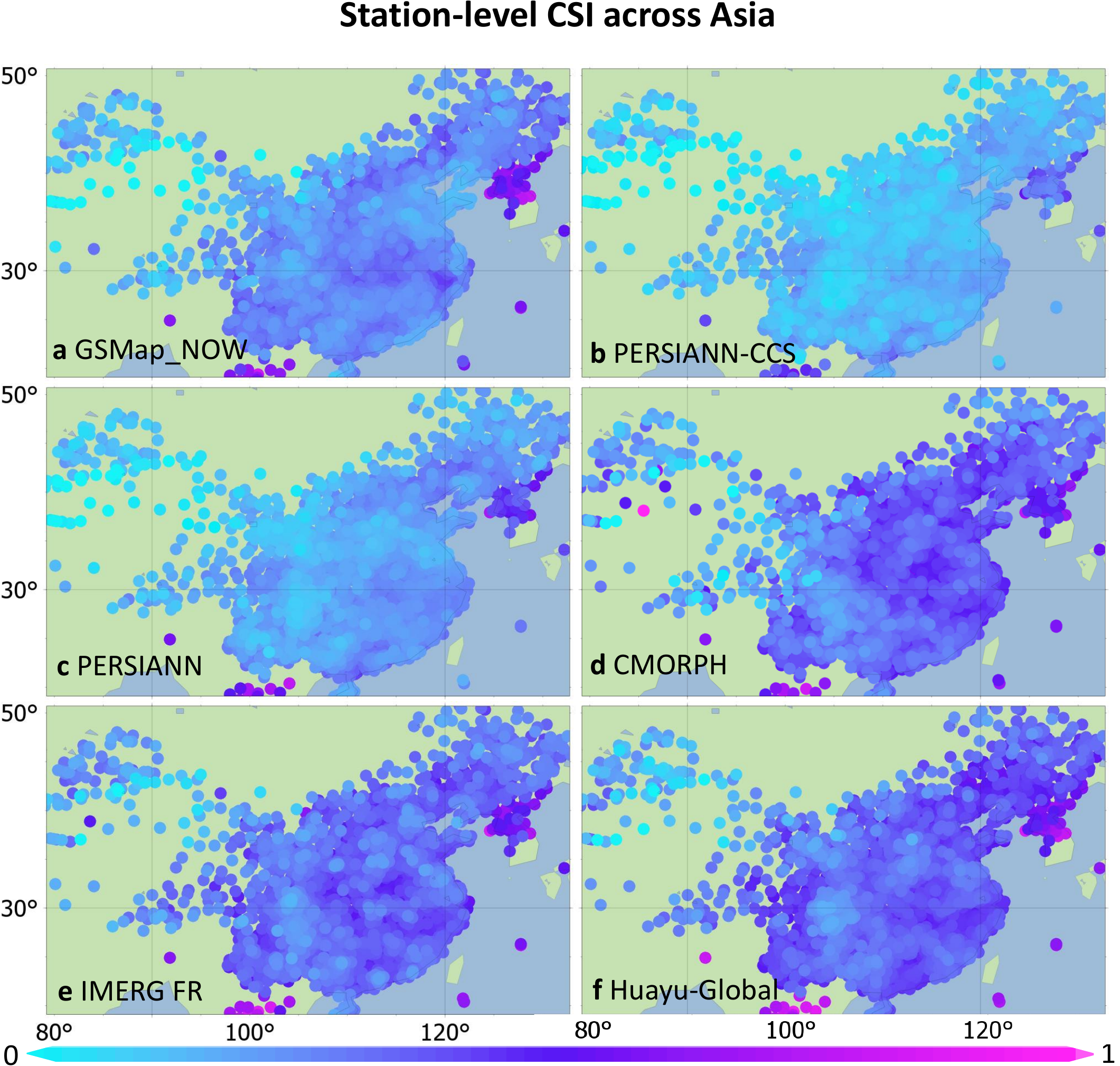}
\caption{The same as Extended Data Fig.~\ref{fig:station_csi_americas} but across Asia.}
\label{fig:station_csi_asia}
\end{figure}

\clearpage

\begin{table*}[!ht]
  \scriptsize
  \setlength{\extrarowheight}{2pt}
  \centering
  \caption{Per-sensor training configuration of Huayu-Global. The three sector models share identical architecture and parameter counts, while learning rate and decay schedule are tuned separately for ABI, AGRI, and SEVIRI streams. All models use AdamW optimization, a batch size of 16, and 100 training epochs on four NVIDIA RTX 6000 Ada GPUs (48~GB each). The implementation uses PyTorch with Python~3.12 on Linux. Training data cover 2023 and 2024; model checkpoints are selected according to held-out validation performance.}
  \label{tab:config}
  \begin{tabular}{
    p{3cm}<{\centering}
    p{2.5cm}<{\centering}
    p{2.5cm}<{\centering}
    p{3cm}<{\centering}
  }
    \toprule
    \tabincell{c}{\textbf{Parameter}}
    & \tabincell{c}{\textbf{Huayu-ABI}}
    & \tabincell{c}{\textbf{Huayu-AGRI}}
    & \tabincell{c}{\textbf{Huayu-SEVIRI}} \\
    \noalign{\vskip 3pt}
    \midrule[0.75pt]
    \tabincell{c}{Satellite} & GOES-16/18/19 & FY-4B & Meteosat-9/10/11 \\
    \noalign{\vskip 3pt}
    \tabincell{c}{Learning Rate} & 0.002 & 0.001 & 0.001 \\
    \noalign{\vskip 3pt}
    \tabincell{c}{Model Parameters} & 241.78M & 241.78M & 241.78M \\
    \noalign{\vskip 3pt}
    \tabincell{c}{LR Decay Schedule} & [10, 20, 30, 40, 50] & [20, 30, 40, 50] & [20, 30, 40, 50] \\
    \noalign{\vskip 3pt}
    \tabincell{c}{Training Samples} & 51,896 & 23,642 & 49,691 \\
    \noalign{\vskip 3pt}
    \tabincell{c}{Checkpoint Epoch} & 93rd & 100th & 99th \\
    \bottomrule
  \end{tabular}
\end{table*}

\begin{table*}[!ht]
\scriptsize
\setlength{\tabcolsep}{4pt}
\centering
\caption{Operational geostationary satellites used in Huayu-Global. The Americas sector is served by GOES-16/18/19 ABI observations, the Europe-Africa sector by Meteosat-9/10/11 SEVIRI observations, and the Asia-Pacific sector by FY-4B AGRI observations. These three systems together cover most low- and mid-latitude regions (60$^\circ$S--60$^\circ$N). GOES-16 and GOES-19 are handled as a continuous ABI stream according to provider operational metadata, with GOES-19 contributing after the GOES-East transition in 2025; Meteosat-10 and Meteosat-11 are likewise handled as a continuous SEVIRI stream. The coverage of AGRI (FY-4B) differs before and after 31~January~2024 owing to a change of the sub-satellite point (from 133$^\circ$E to 105$^\circ$E); the mosaicking procedure accounts for the corresponding sector footprints. The table lists the nominal longitude, launch date, imaging payload, spectral bands, native sampling, and provider for each platform.}
\label{tab:satellite} % 调整标签以区分原表格（可选）
\begin{tabular*}{\textwidth}{@{\extracolsep{\fill}}lcccccc@{}}
\toprule
% 新表头：原卫星型号转为第一列的表头，原参数转为列标题
\tabincell{c}{Satellites}
&\tabincell{c}{Nominal\\longitude(s)}
&\tabincell{c}{Launch date}
&\tabincell{c}{Payloads}
&\tabincell{c}{Bands}
&\tabincell{c}{Native\\resolution}
&\tabincell{c}{Provider}
\\
\noalign{\vskip 3pt}
\midrule[0.75pt]
% 每行对应原表格的一颗卫星，数据按“位置→分辨率→载荷→波段→所属区域”顺序排列
\tabincell{c}{FengYun-4B}
&\tabincell{c}{105°E/133°E}
&\tabincell{c}{2nd June 2021}
&\tabincell{c}{AGRI}
&\tabincell{c}{15}
&\tabincell{c}{4km/15min}
&\tabincell{c}{China}
\\
\noalign{\vskip 3pt}
\tabincell{c}{GOES-16\\GOES-19}
&\tabincell{c}{75.2°W}
&\tabincell{c}{19th Nov 2016\\25th Jun 2024}
&\tabincell{c}{ABI}
&\tabincell{c}{16}
&\tabincell{c}{2km/10min}
&\tabincell{c}{U.S.A.}
\\
\noalign{\vskip 3pt}
\tabincell{c}{GOES-18}
&\tabincell{c}{137°W}
&\tabincell{c}{1st March 2022}
&\tabincell{c}{ABI}
&\tabincell{c}{16}
&\tabincell{c}{2km/10min}
&\tabincell{c}{U.S.A.}
\\
\noalign{\vskip 3pt}
\tabincell{c}{Meteosat-10\\Meteosat-11}
&\tabincell{c}{0°}
&\tabincell{c}{5th July 2012\\15th July 2015}
&\tabincell{c}{SEVIRI}
&\tabincell{c}{11}
&\tabincell{c}{3km/15min}
&\tabincell{c}{Europe}
\\
\noalign{\vskip 3pt}
\tabincell{c}{Meteosat-9}
&\tabincell{c}{45°E}
&\tabincell{c}{22nd Dec 2005}
&\tabincell{c}{SEVIRI}
&\tabincell{c}{11}
&\tabincell{c}{3km/15min}
&\tabincell{c}{Europe}
\\
\bottomrule
\end{tabular*}
\end{table*}

\begin{table*}[!t]
\scriptsize
\setlength{\extrarowheight}{2pt}
\setlength{\tabcolsep}{4pt}
\centering
\caption{Data inventory used in this study. The table summarizes raw geostationary observations, baseline precipitation products, independent station validation data, and processed training tensors, with a total rounded volume of 59.7~TB. Processed ABI, SEVIRI, and AGRI tensors are generated by collocating geostationary imagery with IMERG FR reference fields for the three Huayu-Global sector models. Baseline products span real-time (GSMaP\_NOW, PERSIANN-CCS), short-delay (IMERG ER, IMERG LR), and delayed post-processed (PERSIANN, CMORPH, IMERG FR) streams. Ground validation uses 5,395 HadISD and CMA stations contributing 1,847,967 three-hourly records from two validation windows (July--December~2022 and January--June~2025) completely outside the 2023--2024 training period.}
\label{tab:datasize}
\begin{tabular*}{\textwidth}{@{\extracolsep{\fill}}lccccc@{}}
\toprule
Name & Data Size & Format & Provider & Type & Covered Region \\
\midrule
GOES-16 & 9.2 T & grid & NOAA & Full disk & The Americas \\
GOES-18 & 11 T & grid & NOAA & Full disk & The Americas \\
GOES-19 & 2.4 T & grid & NOAA & Full disk & The Americas \\
Meteosat-10/11 & 14 T & grid & EUMETSAT & Full disk & {Europe-Africa} \\
Meteosat-9 & 5.7 T & grid & EUMETSAT & Full disk & {Europe-Africa} \\
FY-4B & 12 T & grid & NSMC & Full disk & {Asia-Pacific} \\
\midrule
GSMaP\_NOW & 77 G & grid & JAXA & Precipitation & Global \\
PERSIANN & 6.4 G & grid & CHRS & Precipitation & Global \\
PERSIANN-CCS & 23 G & grid & CHRS & Precipitation & Global \\
CMORPH & 15 G & grid & NOAA & Precipitation & Global \\
IMERG ER & 478 G & grid & NASA \& JAXA & Precipitation & Global \\
IMERG LR & 478 G & grid & NASA \& JAXA & Precipitation & Global \\
IMERG FR & 478 G & grid & NASA \& JAXA & Precipitation & Global \\
\midrule
HadISD & 31 G & point & Met Office & Precipitation & Global \\
CMA-station & 603 M & point & CMA & Precipitation & China \\
\midrule
Processed ABI & 2.1 T & grid & - & Tensor & - \\
Processed SEVIRI & 1.1 T & grid & - & Tensor & - \\
Processed AGRI & 611 G & grid & - & Tensor & - \\
\midrule
\textbf{Total} & 59.7 T & - & - & - & - \\
\bottomrule
\end{tabular*}
\end{table*}

\end{document}